\documentclass[preprint,12pt]{elsarticle}

\usepackage{amssymb}
\usepackage{graphicx}
\usepackage{dcolumn}
\usepackage{bm}
\usepackage{amssymb}
\usepackage{epsfig}
\usepackage[utf8]{inputenc} 
\usepackage{ucs} 
\usepackage{amsmath} 
\usepackage{array}
\usepackage{float}
\usepackage{color}
\usepackage[usenames,dvipsnames]{xcolor}
\usepackage{epstopdf}
\usepackage{natbib}
\usepackage{lineno}
\usepackage{color}
\usepackage{subcaption}
\usepackage{caption}
\usepackage{hyperref}
\definecolor{red}{rgb}{1.0,0.0,0.0}
\definecolor{blue}{rgb}{0.0,0.0,1}

\newcommand{\be}{\begin{equation}}
\newcommand{\ee}{\end{equation}}
\newcommand{\bea}{\begin{eqnarray}}
\newcommand{\eea}{\end{eqnarray}}
\newcommand{\beas}{\begin{eqnarray*}}
\newcommand{\eeas}{\end{eqnarray*}}

\newcommand{\nn}{\nonumber}

\journal{Web of Conferences Journal}
\begin{document}
\begin{frontmatter}
\title{Using the Linear Sigma Model with quarks to describe the QCD phase diagram and to locate 
the critical end point}
%
%
\author[a,b]{Alejandro Ayala}
\author[a]{Jorge David Castaño-Yepes}
\author{Jos\'e Antonio Flores\corref{cor1}\fnref{a}}
\ead{jose.flores@correo.nucleares.unam.mx}
\author[a]{Sa\'ul Hern\'andez }
\author[a]{Luis Hern\'andez}

 \address[a]{Instituto de Ciencias Nucleares, Universidad Nacional Aut\'onoma de M\'exico, M\'exico Distrito Federal, C. P. 04510, M\'exico}
 
  \address[b]{Centre for Theoretical and Mathematical Physics, and Department of Physics, University of Cape Town, Rondebosch 7700, South Africa.}

\begin{abstract}
    We study the QCD phase diagram using the linear sigma model coupled to quarks. We compute the effective potential at finite temperature and quark chemical potential up to ring diagrams contribution. We show that, provided the values for the pseudo-critical temperature $T_c = 155$ MeV and critical baryon chemical potential $\mu_{Bc} \simeq 1$ GeV, together with the vacuum sigma and pion masses. The model couplings can be fixed and that these in turn help to locate the region where the crossover transition line becomes first order.
\end{abstract}
\end{frontmatter}

%
%
 
\section{Introduction}
\label{intro}
The description of the QCD phase diagram on the $T$ and  $\mu$ plane reveals profound information for the different phases of strongly interacting matter under extreme conditions such that high temperatures and densities. Most of our knowledge of this phase diagram  is restricted to the region for low values of $\mu$. Lattice QCD has found values for a crossover transition with a critical temperature $T_{c} \sim 155$ MeV  considering $2+1$ quark flavours Ref.~\cite{RefJ1}. On the other hand, effective models  find that for $T \sim 0$ there is a first order phase transition Ref.~\cite{RefJ2}. This means that there must be a point in the diagram where both transitions converge and such point is generally refered to  as the critical end point (CEP). In this work we used the Linear Sigma Model coupled to quarks (LSMq) to locate the CEP. We organize the content as follows: In Sec. 2, we give an overview of the main the properties of LSMq. In Sec. 3, we show the effective potential at high and low-temperature. In Sec. 4, we use the effective potential to determine the coupling constants and to locate the CEP. Finally we summarize and conclude in Sec 5.

\section{Linear Sigma Model coupled to quarks}

We study the restoration of the chiral symmetry using an effective model that accounts for the physics of the spontaneous symmetry breaking at finite temperature and density, the linear sigma model. In order to account to the fermion degrees of freedom around the phase transition, we also include quarks in this model. The Lagrangian for this model is given by 

\bea
\mathcal{L}&=& \frac{1}{2}(\partial_{\mu} \sigma)^{2} +\frac{1}{2}(\partial_\mu \vec{\pi})^{2} + \frac{a^{2}}{2}(\sigma^{2} + \vec{\pi}^{2}) - \frac{\lambda}{4}(\sigma^{2} +  \vec{\pi}^{2})^{2}+ i \overline\psi \gamma ^{\mu} \partial _{\mu} \psi\nn\\
&-&   g\overline \psi (\sigma +i\gamma_{5}\vec{\tau}\cdot \vec{\pi})\psi
\eea            
where  $\psi$ is an SU(2) isospin doublet, $\sigma$ is an isospin singlet and $\vec{\pi}$ is an isospin triplet. $\lambda$ is the boson's self-coupling and $g$ is the fermion-boson coupling. $a^{2} > 0 $ is the mass parameter. 
The Lagrangian admits a broken symmetry vaccum solution given by the minimum of the classical potential when $\lambda$ is positive, this means that the  
sigma field $\sigma$ develops a vacuum expectation value $v$ that becomes in the order parameter of the theory. Shifting the field as $\sigma + v_{0}$, notice the three pions, sigma and the constituent quarks develop the vacuum masses 
\bea
m_{\sigma}^{2} = 3\lambda v^{2}- a^{2}, \;\;  m_{\pi}^{2} = \lambda v^{2}-a^{2},\;\; m_{f} = gv,
\eea
respectively. We take their conservative vacuum values as $m_{\sigma} = 450$ MeV, $m_{\pi} = 140$ MeV,  $m_{f}=300$ MeV Ref.~\cite{RefJ}. Finally, we can fix the value of $a$ using these vacuum values togheter with the first of Eqs. (2)

\begin{equation}
    a = \sqrt{\frac{m_{\sigma}^2 - 3 m_{\pi}^2}{2}}.
\end{equation}

\section{Effective potential}
\label{sec-1}
 We compute the effective potential and the self-energies in the limit where the masses and quark chemical potential are small compared to the temperature, and we include contributions up to ring diagrams. Taking the renormalization scale as $\tilde{\mu} = ae^{-1/2}$, see Ref.~\cite{RefJ3}. For low temperatures, we compute up to the 1-loop correction which is given by the expressions in Ref.~\cite{RefJ4}. To compute 1-loop contribution within both approximations, we start from the following expressions

  \bea
   V_{\text{boson}}&=&T\sum_{i=\sigma,\vec{\pi}}\sum_{n=-\inf}^{\inf}\int\frac{d^3k}{(2\pi)^3}\ln\left(\frac{1}{k^2+m_i^2+\omega_n^2}\right),\nn\\
   V_{\text{fermion}}&=&T\sum_{i=\sigma,\vec{\pi}}\sum_{n=-\inf}^{\inf}\int\frac{d^3k}{(2\pi)^3}\ln\left(\frac{1}{k^2+m_i^2+\left(\tilde{\omega}_n-i\mu\right)^2}\right),
   \eea
 where $\omega_{n} = 2n\pi T$ and $\tilde{\omega}_{n} = (2n+1)\pi T$.


 \subsection{Effective potential at high temperature}
 \label{sec-2}
   The effective potential in the high temperature approximation is computed in Ref.  [3] and is given explicitly by

         \bea V^{h}_{eff} &=& -\frac{1}{2}a^{2}v^{2}+\frac{1}{4}\lambda v^{4}\nn\\
         &-&\sum_{i =\sigma, \pi}\left\{\frac{m_{i}^{4}}{64\pi ^{2}} \left[\ln\left(\frac{a^{2}}{4\pi T^{2}}\right)+\frac{1}{2}-\gamma_{E} \right]\right\}+ N_{c}N_{f}\frac{(gv)^{4}}{16\pi ^{2}}\left[ \ln\left(\frac{4\pi ^{2}a^{2}}{(gv)^{2}}\right) +\frac{1}{2} -\gamma_{E} \right]\nn\\
         &+& \sum_{i = \sigma, \pi}\left\{  \frac{T^{2} m_{i}^{2}}{24} - \frac{\pi ^{2}T^{2}}{90} - \frac{T}{12 \pi}\big [(m_{i})^{2} +  \Pi(T,\mu)\big ]^{3/2} \right\}\\
         &-& N_{c} N_{f}\frac{T}{\pi ^{2}}\int_{0}^{\infty} k^{2}\left[\ln\left(1 + \exp\left(- \frac{\sqrt{ k^{2} + (gv)^{2}} - \mu }{T}\right)\right)\right.\nn\\
         &+&\left.\ln\left(1 + \exp\left(- \frac{\sqrt{ k^{2} + (gv)^{2}} + \mu }{T}\right)\right)\right]\nn
         \eea

             where the self-energy has the expression 

       \begin{equation}
           \Pi[T,\mu] = \underbrace{\frac{\lambda T^{2}}{2}}_{bosons}   \underbrace{-\frac{N_{c}N_{f}g^{2}T^{2}}{\pi ^{2}}\bigg [ Li_{2}(-e^{\frac{\mu}{T}}) + Li_{2}(-e^{\frac{-\mu}{T}}) \bigg ]}_{fermions}. 
       \end{equation}


\subsection{Effective potential at low temperature}
 \label{sec-3}
The effective potential in the low temperature approximation is calculated in the same fashion as Ref.~\cite{RefJ3} and the result is
  
        \bea
        V^{l}_{eff}&=&-\frac{1}{2}a^{2}v^{2} + \frac{1}{4}\lambda v^{4} -\sum _{i = \sigma, \pi}\left \{ \frac{m_{i}^{4}}{64} \left[\ln\left(\frac{4\pi a^{2}}{(b\mu) + \sqrt{(b\mu)^{2} - m_{i}^{2} }}\right) -\gamma_{E} + \frac{1}{2}\right]\right.\nn\\
        &-&\left.\frac{1}{96 \pi ^{2}} (b\mu) \sqrt{(b\mu)^{2} - m_{i}^{2}}- \frac{7 \pi ^{2} T^{4}}{180}\frac{\mu (2(b\mu)^{2}-5m_{i})}{((b\mu)^{2} - m_{i}^{2})^{3/2}} \right\}\nn\\
        &+& N_{f} N_{c}\left [\ln\left( \frac{4\pi ^{2}a^{2}}{(\sqrt{\mu ^{2} -(gv)^{2}} + \mu)^{2} }\right ) + \frac{1}{2} -\gamma _{E} \right]\nn\\
        &+& N_{f} N_{c}\left\{ \frac{\mu \sqrt{\mu^{2}-(gv)^{2}}(2\mu^{2} - 5(gv)^{2} )}{24\pi ^{2}}\right.\nn\\
        &+&\left. \frac{1}{6}T^{2}\mu\sqrt{\mu ^{2} - (gv)^{2}} + \frac{7\pi T^{2}}{360}\frac{\mu(2\mu^{2}) - 5(gv)^{2}}{[\mu ^{2} -(gv)^{2}]^{3/2}}\right \},
        \eea

   where the boson chemical potential $\mu_{b}$ is taken as a fraction $b$ of the quark chemical potential, namelly, $b = \mu_{b}/\mu_{q}$ and the baryon chemical potential is given by $\mu_{B} = 3 \mu _{q}$. The boson chemical potential $\mu_{b}$  provides the information on the average number of boson particles interacting between both phases at high densities.

\section{Coupling Constants and location of CEP}

\begin{figure}[h]
\includegraphics[scale=0.5]{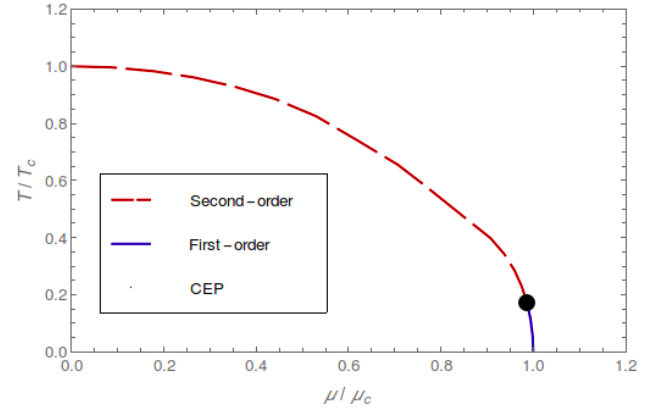}
\centering
\caption{Effective QCD phase diagram obtained for $\lambda = 0.897$ and $g = 1.57$, taking the critical temperature for two light flavors  as $T_{c} = 170$ MeV at $\mu = 0$ MeV and critial quark chemical potential $\mu_{qc} = 340$ MeV at $T = 0$ MeV. The dashed line represents the second order transition and the solid line represents the first order transition. The CEP is located at $(\mu^{CEP}/T_{c},T^{CEP}/{T_{c}}) \sim (0.993,0.113)$.}
\label{fig-for-reference}
\end{figure} 

To determine a unique transition curve of the QCD phase diagram, we compute the coupling constants and the critical values $(\mu _{Bc},T_{c})$, in order to determine the parameters $\lambda$ and $g$, we solve a system of equations obtained from the conditions




	

\begin{equation}
\frac{dV^{l}_{eff}}{dv}\bigg|_{v=0,T=0,\mu = \mu_{c} } = \frac{dV^{l}_{eff}}{dv}\bigg|_{v=v_{1},T=0,\mu = \mu_{c} } = 0
\end{equation}

\begin{equation}
    V^{l}_{eff}(0,0,\mu_{c}) = V^{l}_{eff}(v_{1},0,\mu_{c}) 
\end{equation}


\begin{equation}
     [m_{\pi}^{2}(v) + \Pi(T,\mu) ]\bigg|_{v=0,T=T_{c},\mu = 0} = [m_{\pi}^{2}(v) + \Pi(T,\mu) ]\bigg|_{v=(m_{q}/3),T=0,\mu = \mu_{c} } = 0
\end{equation}

where the critical temperature is taken as $T_{c} = 170$ Mev for 2 light flavors, see Ref.~\cite{RefJ6}, the critical quark chemical potential $\mu_{c} \sim 340$ MeV Ref.~\cite{Hagedorn}, the dynamical quark mass $m_{q} = 300$ MeV and the $v_{0}$ is the expectation value of sigma. Notice that on the boundary of the first order phase transitions the effective potential shows two minima, one at $v = 0$ and the other at $v = v_{1}$. By solving the Eqs. (9), (10) and (11) the coupling constants are $\lambda = 0.897$ and $g = 1.57$. For details see Ref.~\cite{RefJ7}.
Finally, we consider an interpolation between both approximations of the effective potential finding the lines that correspond to the phase order transitions. The region where they converge locates the position of the CEP which is approximately given by $(\mu^{CEP}/T_{c},T^{CEP}/{T_{c}}) \sim (0.993,0.113)$.



\section{Conclusion}
 In this work we have explored the QCD phase diagram using the LSMq considering the approximation of the effective potential in the high temperature up to ring diagrams order and in the low temperature working up to 1-loop correction. The phase diagram derived within these approximation gives us information on the location of the CEP and we conclude that the LSMq is a suitable effective model to describe the phase transition in the temperature and density plane to understand much better the chiral restoration symmetry.

 



\begin{thebibliography}{}
%
%
\bibitem{RefJ1}
T. Bhattacharya, \textit{et al.}, Phys. Rev. Lett. \textbf{113}, 082001 (2014).

\bibitem{RefJ2}
M. Asakawa and K. Yazaki, Nucl. Phys. A \textbf{504}, 668 (1989); A. Barducci, R. Casalbuoni, S. De curtis, R. Gato and G. Pettini, Phys. Lett. B \textbf{231}, 463 (1989); Phys. Rev. D \textbf{41}, 1610 (1990). 


\bibitem{RefJ}
C. Pratignani, \textit{et al.} (Particle Data Group), Chin. Phys. \textbf{C} 40, 100001 (2016).


\bibitem{RefJ3}
C. O. Dib, O. R. Espinosa, Nucl. Phys. B \textbf{612}, 492-518 (2001).


\bibitem{RefJ4}
M. Le Bellac, \textit{Thermal Field Theory} (Cambridge University Press, Cambridge, 1996).


\bibitem{RefJ5}
A. Ayala, C. A. Bashir, J. J. Cobos-Martinez, S. Hernandez-Ortiz, A. Raya, Nucl. Phys. B \textbf{897}, 77-86 (2015).


\bibitem{RefJ6}
 Y. Maezawa, S. Aoki, S. Ejiri, T. Hatsuda, N. Ishii, K. Kan
aya, N. Ukita, J. Phys. G  \textbf{34}, S651 (2007).


\bibitem{Hagedorn}
R. Hagedorn, Nuovo Cim. Suppl. \textbf{3} (1965) 147; Nuovo Cim. A \textbf{56} (1968) 1027. 

\bibitem{RefJ7}
A. Ayala, L. Hern\'andez, S. Hern\'andez Ort\'iz, e-print arXiv: 1710.09007 [hep-ph].






 




\end{thebibliography}
%
%

\end{document}